\documentclass[%
 reprint,
 amsmath,amssymb,
 aps,
]{revtex4-2}
\usepackage{amsmath}
\usepackage{amsthm} 
\usepackage{graphicx}
\usepackage{dcolumn}
\usepackage{physics}
\usepackage{bm}
\usepackage{color}

\usepackage{braket}
\usepackage{ragged2e}

\theoremstyle{plain}

\theoremstyle{definition}
\newtheorem{dfn}{Definition}
\newtheorem{prop}{Proposition}

\begin{document}

\preprint{APS/123-QED}

\title{Private quantum network sensing with efficient multi-partite entanglement\\ distribution via lossy channels}

\author{Yoshihiro Ueda$^{1}$}
 \email{$\rm yoshihiro_u@keio.jp$}

\author{Makoto Ishihara$^{1}$}%

\author{Wojciech Roga$^{1}$}%

\author{Masahiro Takeoka$^{1,2}$}%
\affiliation{%
$^{1}$Department of Electronics and Electrical Engineering, Keio University, 3-14-1 Hiyoshi, Kohoku-ku, Yokohama 223-8522, Japan} 

\affiliation{%
$^{2}$Advanced ICT Research Institute, National Institute of Information and Communications Technology (NICT), Koganei, Tokyo 184-8795, Japan
}

\date{\today}

\begin{abstract}
Quantum network sensing protocols show potential to enhance the estimation precision for functions of spatially distributed parameters beyond the shot noise limit. Such protocols can also provide privacy, preventing sensitive information from leaking to unauthorized parties. The key resource required for these tasks is multipartite quantum entanglement. The photonic entanglement is the most natural in this context, however, distributing it over long distances presents significant difficulties, mainly because of unavoidable loss in communication channels. 
The resource efficiency is also fundamental, both for precise network sensing and for private sensing. In this research, we analyze quantum network sensing protocols based on a recently proposed, efficient Greenberger-Horne-Zeilinger (GHZ) state distribution scheme. We 
compare our protocol with conventional methods based on direct entanglement distribution to show that it
reduces the loss-induced estimation error of certain functions of distributed parameters. 
Moreover, we consider a scenario in which one person
can estimate linear combinations of distributed parameters without violating privacy of the other users. 
We show that an additional parameter 
can be used to hide the information about the target parameter from everyone, except the person who controls the parameter.
\end{abstract}

\maketitle


\section{\label{sec:level1}Introduction}

Entanglement-based quantum sensing \cite{qs,qes,apqs} is known as a sensing method which can surpass the classical limit in terms of measurement precision. Specifically, measuring a parameter $\theta$ repeatedly with $\textit{N}$ classical probes, such as light pulses, results in the estimation error scaling according to the so-called standard quantum limit (SQL), 1/$\sqrt{\textit{N}}$. In contrast, by using quantum entangled probes $\textit{N}$ times, the estimation error scaling becomes 1/$\textit{N}$, referred to as the Heisenberg limit (HL) \cite{HL,HL2}. This can appear advantageous in various applications ranging from microscopy \cite{quantum_microscope} to gravitational wave detection \cite{LIGO} in which the number of probes is naturally limited.
Moreover, quantum-enhanced network sensing aiming at a precise estimation of multiple parameters and their linear combinations can be applicable in quantum network scenarios \cite{dqs1,dqs2,dqs3} such as precise clock synchronization \cite{clock_sync1,clock_sync2} and the long-baseline quantum telescope \cite{QT_g,QT_CV,QT_DV}.

Recent results show that both the continuous-variable (CV) \cite{dqs_cv1,dqs_cv2, dqs_cv3} and discrete-variable (DV) \cite{dqs_dv1,dqs_dv2,dqs_dv3} multipartite entanglement can be useful in enhanced-precision phase estimation with respect to the classical methods. In particular, Greenberger-Horne-Zeilinger (GHZ) states \cite{GHZ0} are known to be the optimal probes which can attain the HL \cite{qes}. 
Quantum advantages based on CV  and DV \cite{dqs_dv1exp,dqs_dv2exp,dqs_dv3exp,dqs_dv4exp} entanglement was also observed experimentally, for example in refs. \cite{dqs_cv1exp,dqs_cv2exp} and \cite{dqs_dv1exp,dqs_dv2exp,dqs_dv3exp,dqs_dv4exp}, respectively.

Quantum network sensing has potential not only to enhance the estimation precision but also to hide the information about local parameters~\cite{SecureQNSTheo,SecureQNSExp,SecureQNSMulti}. 
It allows us to estimate unknown parameters without revealing them to the others. Recently, two-party implementation has been proposed~\cite{SecureQNSTheo} and experimentally demonstrated~\cite{SecureQNSExp}. It aims at sending confidential information, such as medical data of patients guaranteeing privacy and security. The protocols in refs.~\cite{SecureQNSTheo,SecureQNSExp} were inspired by quantum teleportation and blind quantum computing. They succeeded in estimating a parameter while hiding the information from a potential eavesdropper. Furthermore, a multi-user protocol has also been proposed~\cite{SecureQNSMulti}. In that scenario, the users can know a linear combination of the parameters and their own local parameter, but cannot know the local parameters of the other users. 
The GHZ state \cite{qpds} and the continuous variable entangled \cite{PrivacyCV} states can be used as a resource. 

\begin{figure}[htbt]
    \centering
    \includegraphics[width=1\linewidth]{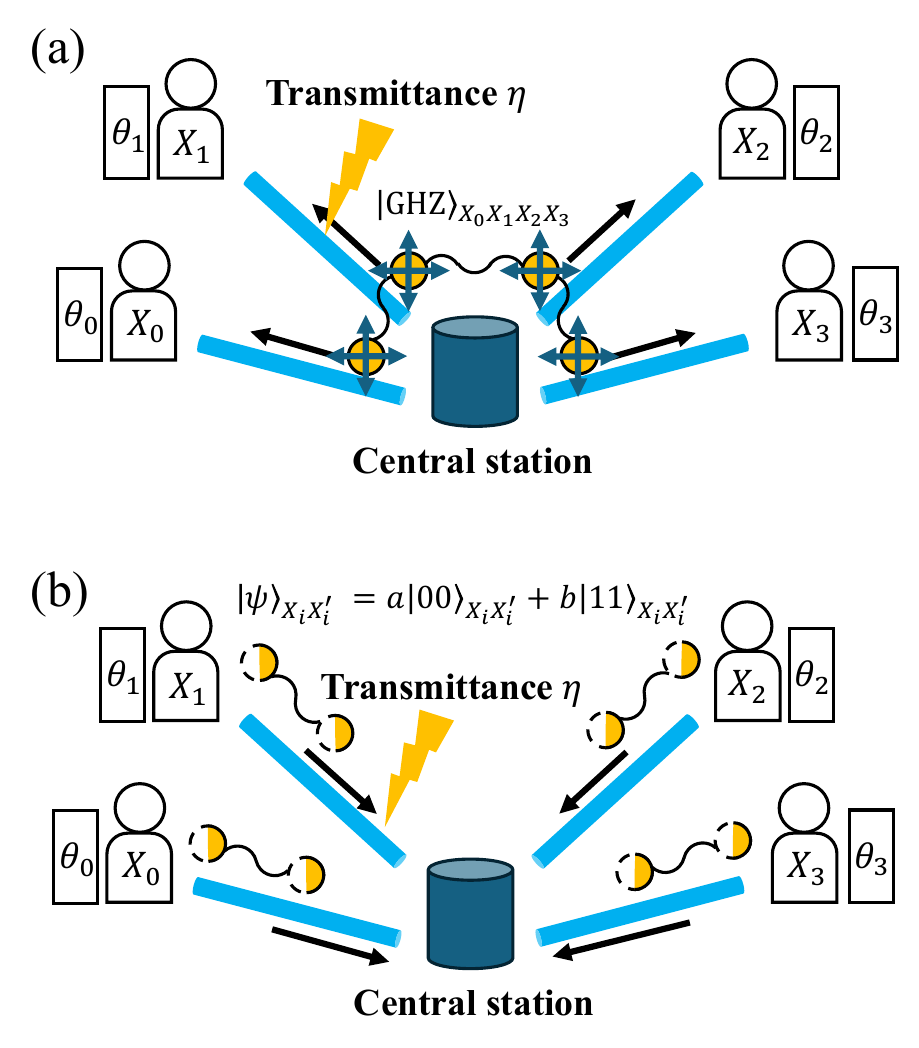}
    \caption{A star network for estimation of functions of target parameters $\theta_0$,...,$\theta_3$ using distributed 4 partite entangled state $|\psi\rangle_{X_0X_1X_2X_3}$. The stations are denoted by $X_0$,...,$X_3$.  
    The sensing stations are connected with a central station with equal length optical links with transmittance $\eta$. (a) Direct transmission protocol. A polarized GHZ state is generated in the central station and is transmitted to the sensing stations via lossy channels. (b) Our scheme. Each station prepares a bipartite photonic state $\ket{\psi}_{X_{i}X_i^{'}}$ and sends one part to the central station, where the photons interfere and their states are measured in the photon number basis.}
    \label{1}
\end{figure}

A typical optical quantum network sensing setup consists of several sensing stations connected in a network via optical fibers characterized by a given transmittance per km, which is illustrated in Fig.~\ref{1}(a). 
Photonic entangled states are generated at the central station and are sent to the distant parties to estimate the local parameters, $\theta_i$. 
To leverage the advantage of entanglement, all photons must be successfully transmitted. If the transmission rate is $\eta$, the probability of this process scales as $\mathcal{O}(\eta^M)$ where $M$ is the number of nodes. It is especially critical for long-distance transmission, where $\eta\ll 1$, in which an effective precision of the parameter estimation is severely limited. 
We call this as the direct transmission scenario. 
It suggests an importance of entanglement distribution protocols that are more robust against channel losses. 
Although several works have investigated the performance of quantum network sensing, including the effect of noisy states and environment \cite{dqs_noise1,dqs_noise2,dqs_noise3}, the entanglement distribution process with long-distance lossy communication channels is rarely included in the analysis. 
The channel losses may be a further problem in the private network sensing scenario. 
The recent paper~\cite{Private_Robust_QNS} revealed that some types of noise disrupt the privacy. Thus, it is important to clarify if the losses also disrupt the parivacy or not. 

In this paper, we propose a private distributed sensing protocol based on a loss tolerant GHZ state distribution protocol proposed in \cite{shimizughz}. 
The schematic of our protocol is illustrated in Fig.~\ref{1}(b). 
In our protocol, the local users send a part of the bipartite entangled photons to the central statation in which the photons are interefered and then detected. 
With proper photon detection events, the local users can hold entangled photons. 
We show that our protocol can estimate not only the local parameters, but also a set of linear combinations of the local parameters.
The estimation precision is evaluated by the Fisher information, which provides the lower bound for the variance of estimates via the Cramér-Rao bound. We evaluate the precision of the estimation in the presence of loss in the entanglement distribution by calculating the success probability of the process. We observe that our protocol achieves a lower variance than the direct transmission scenario in a lossy environment. 

In addition to the precision, we redanalyze the privacy in our protocol. The aim is that the central station estimates a linear combination of parameters privately. To achieve this goal, we apply the additional phase on one of the transmitted photons at the centeral station of the star network. We found that this protocol can hide the information about the linear combination of parameters. We also show that our distribution protocol preserves privacy even in the presence losses in the star network channels. 

The paper is organized as follows. In Sec~\ref{sec:level2}, we describe the method of distributing the GHZ states and introduce the Fisher information as a measure of distributed sensing accuracy. In Sec.~III 
we demonstrate that our protocol can measure a set of linear combinations of the local unknown parameters more precisely than the conventional method. 
In Sec.~IV, we discuss and analyze the privacy of our scheme.
In Sec.~V, we summarize the results and discuss them in a wider context. 

\section{\label{sec:level2}Methods}

In this section, we define a loss-tolerant protocol for the distribution of GHZ states. We also define the Fisher information which we use to estimate the precision limit for quantum network sensing. 

\begin{figure}
    \centering
    \includegraphics[width=1\linewidth]{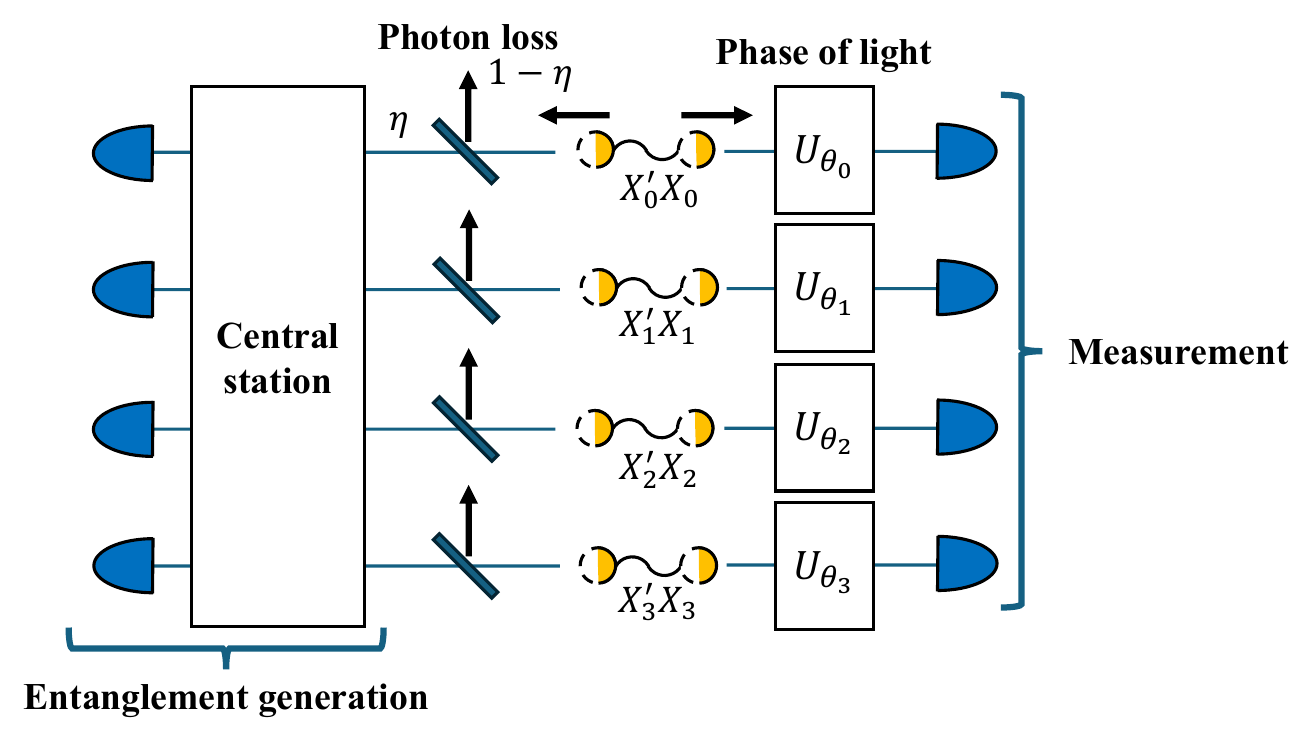}
    \caption{The detail of our scheme. The left side of this figure shows the GHZ state distribution process. The photon loss is modeled by the beam splitters with the transmittance $\eta$. The right side shows the measurement process. We assume the estimated parameters are the phase of light described by $\theta_0,...,\theta_3$.}
    \label{2}
\end{figure}

The GHZ state is a multipartite entangled quantum state widely used in quantum protocols ranging from quantum cryptography to quantum sensing \cite{dqs_dv1, dqs_dv2exp, cka}. 
In the standard basis of two-level quantum systems (qubits) it is defined as a superposition of two perfectly aligned states of all systems. Let us give a definition for 4-qubits, which is relevant to our paper, and the extension of which is straightforward:
\begin{equation}
    \ket{\text{GHZ}}=\frac{1}{\sqrt{2}}(\ket{0000}+\ket{1111}).\label{ghzorg}
\end{equation}
The phase between the terms is a matter of convention about how one defines the basis of particular systems, and does not reflect any essential properties of entanglement. Thus, we can choose the orthogonal basis of any systems, such as the energy levels of atom and the polarization, to describe the GHZ state. Here we mention that the GHZ states in this paper are described as
\begin{equation}
    \ket{\text{GHZ}}=\frac{1}{\sqrt{2}}(\ket{1010}-\ket{0101}).
    \label{ghzdef}
\end{equation}
We can obtain this GHZ state by applying bit-flip and phase-flip to the GHZ state in Eq.~\eqref{ghzorg}. It can be distributed in a network and used to estimate local parameters or their functions.

Let us consider the following GHZ state distribution protocols. We consider a star network shown in Fig.~\ref{1}. The parameters $\theta_0$,...,$\theta_3$ and the respective sensing stations $X_0$,...,$X_3$ are separated from each other. The direct transmission protocol, Fig.~\ref{1}(a), typically applies the polarized GHZ states \cite{dqs_dv2exp,qpds},
\begin{equation}
    \ket{\text{GHZ}}=\frac{1}{\sqrt{2}}(\ket{VHVH}-\ket{HVHV})\label{ghz_polari}
\end{equation}
where $\ket{H}$ and $\ket{V}$ are vertical and horizontal polarization that are orthogonal basis to each other and $\ket{H}=\ket{0}, \ket{V}=\ket{1}$. The GHZ states are generated by the central station locally and distributed to each station via an optical fiber characterized by transmittance $\eta$. In this strategy, it is necessary to distribute all photons to the appropriate stations, therefore, the distribution success probability 
decreases exponentially as $\eta^M$, where $M$ is the number of sensing stations.  

To reduce the effect of loss without using advanced quantum repeaters, an efficient GHZ state distribution scheme was proposed in \cite{shimizughz}. The probability of success scales as $\eta^{\frac{M}{2}}$, which is a quadratic improvement with respect to the direct transmission. This method can be applied to an arbitrary even number of users $\textit{M}$. Here, we consider $\textit{M}=4$ stations which estimate $\textit{d}=4$ local parameters. 

\begin{figure}
    \centering
    \includegraphics[width=1\linewidth]{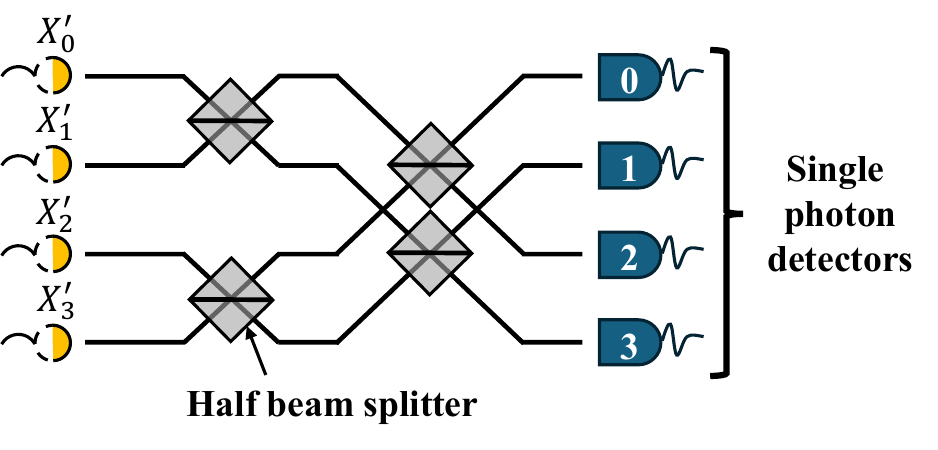}
    \caption{Configuration of central station. Each station sends one of the photon pair to the central station and these transmitted photons are interfered with the interferometer constructed by the half beam splitters and are detected by the single photon detectors. 
    }
    \label{3}
\end{figure}

In our scenario shown in Fig.~\ref{1}(b) and Fig.~\ref{2}, each local sensing station prepares a bipartite state  
\begin{equation}
    {\ket{\psi}}_{X_i X_i^{'}} = a\ket{00}_{X_i X_i^{'}}+b\ket{11}_{X_i X_i^{'}},\label{bell1}
\end{equation}
where $|a|^2+|b|^2=1$. Here, $X_i$ and $X_i^{'}$ denote the subsystems associated with the $i$-th node. We can rewrite the prepared state as $\ket{\Phi}=\left(\bigotimes_{i=0}^3 |\psi\rangle_{X_i X_i'}\right)$. The setup of the central station is shown in Fig.~\ref{3}. the state in Eq.~(\ref{bell1}) can be generated in a photonic or atomic system. In our case, the qubit in the photon-number basis $\ket{0}_{X_i^{'}}$ and $\ket{1}_{X_i^{'}}$ is transmitted to the central station. At the central station, the transmitted qubits interfere with each other and the output state is measured in the photon-number basis. Each observation in the measurement of subsystems $X_0^{'}X_1^{'}X_2^{'}X_3^{'}$ results in collapsing the system $X_0X_1X_2X_3$ to some state. The setup from Fig.~\ref{3} was used before to generate the W and Dicke states \cite{wojciechWDicke}. A few measurement results lead to formation of a GHZ state in the sensing stations \cite{shimizughz}.

To illustrate this mechanism, let us consider an example starting with the ideal lossless case and noiseless detectors. If two detectors register a detection, we know that only two photons arrive at this stage; however, we do not know which systems generated those photons. Therefore, the resulting state, after the measurement, is a superposition of possible scenarios. For instance, if the first and second detectors register photons, we succeed in generating the following GHZ state in photon-number basis
\begin{equation}
    \ket{\text{GHZ}}_4=\frac{1}{\sqrt{2}}\left(\ket{1010}_{X_0X_1X_2X_3}-\ket{0101}_{X_0X_1X_2X_3}\right).\label{GHZ1}
\end{equation}
Depending on the detection pattern, we can also generate different states which, up to local basis renaming or a local phase flip, are equivalent to the GHZ states. These states are listed in Table \ref{t1}. We choose one of three kinds of the GHZ states for the calculation, and we note that the other GHZ states also can be used to estimate the parameters listed in Table \ref{t2}.

In general, considering possibility of losses, after a successful detection event, the generated state can be described as a mixed state
    \begin{equation}
    \rho_{\text{CS}}=p\ketbra{\text{GHZ}}{\text{GHZ}}_4+\sum_{i}r_{i}\ketbra{\phi_i}{\phi_i},\label{GHZCS}
    \end{equation}
where $\ket{\phi_i}$ are some undesirable states that are orthogonal to $\ket{\text{GHZ}}$. Here $p+\Sigma r_i=1$. \\

In the context of quantum network sensing, the distributed state that can be generated in this way can be used to estimate unknown distributed parameters $\theta_i$, where $i=0,...,3$. We assume that the unknown parameters are local perturbations of phases of light, which are introduced to photonic states by unitary transformations.
In this analysis, without loss of generality, we assume that we distribute state $(\ket{1010}-\ket{0101})/\sqrt{2}$. We want to estimate the target parameter,
\begin{eqnarray}
    \theta=\theta_0-\theta_1+\theta_2-\theta_3.    
\end{eqnarray}
After appropriate pattern of detectors registers photons, the final state is as follows:
\begin{eqnarray}
        \rho_{\text{CS},\theta}&=&U_{\theta}\rho_{\text{CS}}U_{\theta}^{\dagger}\nonumber\\
        &=&pU_{\theta}\ketbra{\text{GHZ}}{\text{GHZ}}_{4}U^{\dagger}_{\theta}+\sum_{i}r_i\ketbra{\phi_i}{\phi_i}\label{PGHZstate}
\end{eqnarray}
Here $U_{\theta}$ is the unitary operator representing a phase shift, which acting in the photon-number basis $\{|0\rangle, \ket{1}\}$ is given as
\begin{eqnarray}
    U_{\theta}=U_{\theta_0}\otimes{U}_{\theta_1}\otimes{U}_{\theta_2}\otimes{U}_{\theta_3}, \quad U_{\theta_i}=
    \begin{bmatrix}
        1 & 0 \\
        0 & e^{i\theta_{i}} \\
    \end{bmatrix}
\end{eqnarray}
The undesirable states $\ket{\phi_i}\bra{\phi_i}$ in Eq.~(\ref{PGHZstate}) are diagonal because they are generated by the photon loss. So, they do not carry any information about $\theta$. We can estimate the lower bound of the variance of the parameter $\Delta\theta^2$ 
by calculating the Fisher information. 


\begin{table}[h]
        \centering
        \caption{States $\ket{\Phi}_{X_0, ..., X_3}$ in subsystems $X_{0},...,X_{3}$ depending on detection pattern in subsystems $X_{0}^{'}, ...,X_{3}^{'}$ \cite{shimizughz}. 
        }
        \begin{tabular}{c|c}
        \hline
         Detection pattern & State$\ket{\Phi}_{X_0, ..., X_3}$ \\
        \hline
        \hline
         0 and 1 & $(\ket{1010}-\ket{0101})/\sqrt{2}$ \\ 
        \hline
         0 and 2 & $(\ket{1100}-\ket{0011})/\sqrt{2}$ \\
        \hline
         0 and 3 & $(\ket{1001}-\ket{0110})/\sqrt{2}$ \\
        \hline
         1, and 2 & $(-\ket{1001}+\ket{0110})/\sqrt{2}$ \\
        \hline
         1 and 3 & $(-\ket{1100}+\ket{0011})/\sqrt{2}$ \\
        \hline
         2 and 3 & $(-\ket{1010}+\ket{0101})/\sqrt{2}$ \\
        \hline
        \end{tabular}
        \label{t1}
\end{table}

\begin{table}[h]
        \centering
        \caption{The phases which our protocol can estimate.}
        \begin{tabular}{c|c}
        \hline
         Detection pattern & The estimated parameters \\
        \hline
        \hline
         Pattern 1: \{1,2\} or\{3,4\} & 
         \begin{tabular}{c}
            $\theta_{P1+}=-\theta_0+\theta_1-\theta_2+\theta_3$ \\ or \\ 
         $\theta_{P1-}=\theta_0-\theta_1+\theta_2-\theta_3$
         \end{tabular} \\ 
        \hline
         Pattern 2: \{1,3\} or \{2,4\} &
         \begin{tabular}{c}
            $\theta_{P2+}=-\theta_0+\theta_1+\theta_2-\theta_3$ \\ or\\$\theta_{P2-}=\theta_0-\theta_1-\theta_2+\theta_3$  
         \end{tabular}\\
        \hline
         Pattern 3: \{1,4\} or \{2,3\} & 
         \begin{tabular}{c}
            $\theta_{P3+}=\theta_0+\theta_1-\theta_2-\theta_3$ \\ or \\ $\theta_{P3-}=-\theta_0-\theta_1+\theta_2+\theta_3$
         \end{tabular}\\
        \hline
        \end{tabular}
        \label{t2}
\end{table}

\subsection{\label{sec:citeref}Parameter estimation limit: classical and quantum Fisher information}

Precision in parameter estimation is quantified by the estimator variance $\Delta\theta^2$, which is derived from measured parameters $\theta_{\text{est}}$ and a set of associated probability function $\bm{P}$. The variance $\Delta\theta^2$ is defined as follows:
\begin{equation}
    \Delta\theta^2 = \ev{\left(\theta_{\text{est}}-\theta\right)^2}.
\end{equation}
From the probability function $P_i\in\bm{P}$, we can construct the 
classical Cramér-Rao bound,
\begin{equation}
    \Delta\theta^2 \geq\frac{1}{N F_{\text{C}}[\bm{P}]},\label{CCRB}
\end{equation}
where $F_{\text{C}}$ is the classical Fisher information (CFI) of $\theta$ and $N$ is the number of probes \cite{HL}. In the single parameter scenario, CFI of $\theta$ is defined as follows:
\begin{equation}
    F_{\text{C}}[\bm{P}]=\sum_i\frac{1}{P_i}\left(\frac{\partial P_i}{\partial \theta}\right)^2.
\end{equation}
In the multiparameter scenario, the role of CFI is played by the Fisher information matrix (CFIM),
\begin{equation}
    (\bm{F}_{\text{C}}[\bm{P}])_{kl}=\sum_i\frac{1}{P_i}\left(\frac{\partial P_i}{\partial \theta_k}\right)\left(\frac{\partial P_i}{\partial \theta_l}\right).\label{definition_of_CFIM}
\end{equation}
Each element of CFIM indicates the amount of information related to the corresponding parameters. For example, $(\bm{F}_{\text{C}}[\bm{P}])_{11}$ is the information of $\theta_1$. By taking the inverse of CFIM, we can find the lower bound on the variance of the corresponding parameter,
\begin{equation}
    \Delta^2\theta_i\geq[\bm{F}_{\text{C}}[\bm{P}]^{-1}]_{ii}.
\end{equation}
Note that the CFIM is not always invertible. By diagonalizing the CFIM, we learn which parameters can be estimated with possibly small uncertainty. Indeed, the eigenvalues of the CFIM indicate the amount of information and its inverse bounds the precision of estimation, while the corresponding eigenvectors indicate the linear combinations of the parameters which are estimated up to that precision. The desired parameter should be one of these eigenvectors. The combinations of the parameters corresponing to the zero eigenvalues have infinite variances.

To calculate the CFI for the parameter $\theta$ encoded in the shared entangled state (\ref{PGHZstate}) and measured by local sensing stations, we need to define a measurement operators applied by each sensing station independently. It is known~\cite{dqs_dv2exp} that in such a case, 
the measurement defined by the projectors on the eigenvectors of the Pauli operator $\sigma_x$ 
\begin{equation}
    \begin{split}
        &\sigma_{x+}=\frac{1}{\sqrt{2}}(\ket{0}+\ket{1}),\\
        &\sigma_{x-}=\frac{1}{\sqrt{2}}(\ket{0}-\ket{1}),\label{sigmaxmeasur}
    \end{split}
\end{equation}
is optimal to achieve the minimum of $1/F_\text{C}[\bm{P}]$.

\begin{figure}[hptp]
    \centering
    \includegraphics[width=0.7\linewidth]{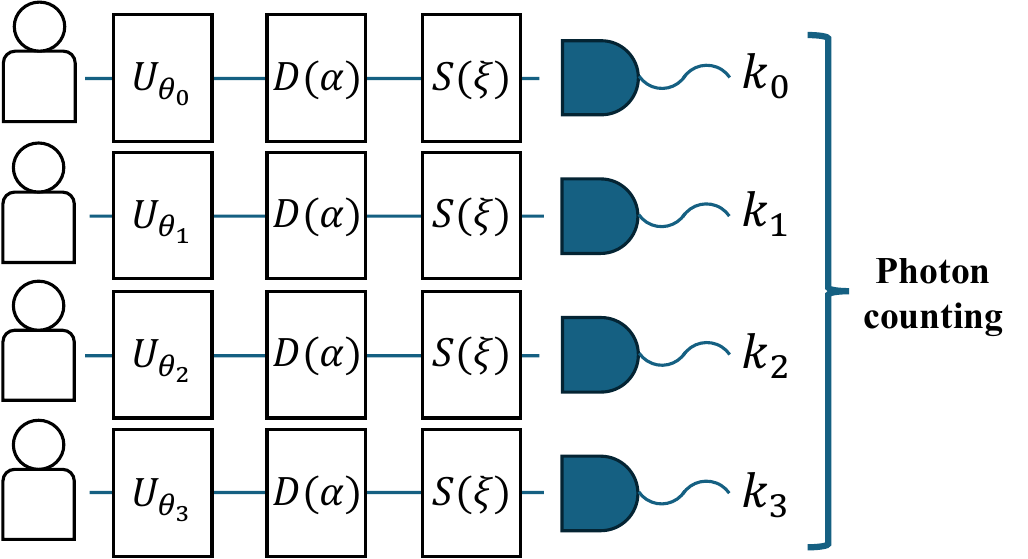}
    \caption{Measurement setup for extracting information about a global parameter $\theta$ from coincidences in the local measurements given by displacement $D\left(\alpha\right)$, squeezing $S\left(\xi\right)$, and photon detection. The measured state is the shared global state. 
    }
    \label{4}
\end{figure}

However, this measurement is particularly problematic in the photon-number qubit basis, as it requires intermediate interaction between the photon and matter, for example, in quantum memory, or other sophisticated techniques including fast feebacks \cite{Takeoka2005, Takeoka2006}. Therefore, instead of $\sigma_x$ basis measurement, we consider the approximated measurement defined by displacement followed by photon detection \cite{takesasa2008}.
In addition, we use squeezing operators to approach the ideal $\sigma_x$ basis measurement. This situation is schematically shown in Fig.~\ref{4}. Thus, our local measurement operators $M_{i,0}$ and $M_{i,1}$ which we propose to extract information about the global parameter $\theta$ from the shared entangled state (\ref{PGHZstate}) are: 

\begin{equation}
    M_{i,0}=\ketbra{\alpha, \xi}{\alpha, \xi},\quad M_{i,1}=I-\ketbra{\alpha, \xi}{\alpha, \xi},\label{displace_and_pc}
\end{equation}
where $I$ is the identity matrix and $\xi=re^{i\varphi}$. Here, $\ket{\alpha, \xi}$ is a squeezed coherent state defined as follows \cite{dss}:
\begin{equation}
    \ket{\alpha, \xi}=\sum_{n=0}^\infty C_n\ket{n},
\end{equation}\begin{equation}
    \begin{split}
        &C_n=N\left[n!\cosh(r)\right]^{-\frac{1}{2}}\left[\frac{1}{2}e^{i\varphi}\tanh(r)\right]^{\frac{n}{2}}\\
        &\hspace{23pt}\times H_n[r(e^{-i\varphi}\sinh(2r))^{-\frac{1}{2}}]
    \end{split}
\end{equation}
where $N=\exp{-\frac{1}{2}\abs{\alpha}^2-\frac{1}{2}(\alpha^{*})^{2}e^{i\varphi}\tanh{r}}$ and $H_n$ are the Hermite polynomials. 
Since each mode contains at most one photon in our protocol, only $\ket{0}$ and $\ket{1}$ components of $\ket{\alpha,\xi}$ contribute. Therefore, we can restrict $\ket{\alpha,\xi}$ to the photon-number qubit subspace as $\ket{\alpha,\xi} \approx C_0\ket{0} + C_1\ket{1}$.

Based on such a measurement, we can construct the probability function $\bm{P}$ for joint detection events of all stations from the four-fold coincidence events. We number these events by $k\in\{0,2^4-1\}$, where each event registers a pattern composed of local measurements $M_{i,k_i}$ and $k_i$ is $0$ or $1$. Thus, $P_k$ can be expressed as
\begin{equation}
P_k=\text{Tr}[\rho_{\theta}M_{0,k_0}\otimes M_{1,k_1}\otimes M_{2,k_2}\otimes M_{3,k_3}].\label{binary_res}
\end{equation}
The index $k\in\{0,2^4-1\}$ in its binary representation is $(k_0k_1k_2k_3)$ and $\rho_{\theta}$ is given by Eq.~(\ref{PGHZstate}). 

The CFI specifies the lower bound on the variance of the estimated parameter. However, the value of this bound depends on how the parameter is measured and we cannot know the tight lower bound. In addition, the optimal measurement for photon-number qubit in terms of current feasible technology is unknown. On the other hand, the quantum Fisher information (QFI) and the quantum Cramér-Rao bound (QCRB) give the lower bound on the CFI and the CCRB and the lower bound on the variance of parameters encoded in quantum states with non-local quantum correlations. Similar to Eq.~(\ref{CCRB}), 
QCRB in the single parameter scenario is given by
\begin{equation}
    \Delta\theta^2\geq\frac{1}{N F_{\text{C}}[\bm{P}]} \geq\frac{1}{N F_{\text{Q}}[\rho_{\theta}]},\label{QCRB}
\end{equation}
where $F_{\text{Q}}$ is the QFI of $\theta$ and $\rho_\theta$ is the density matrix parametrized by $\theta$. 
$F_{\text{Q}}$ is defined as follows:
\begin{eqnarray}
    F_{\text{Q}}=\Tr[\rho_{\theta}L_{\theta}^2]
\end{eqnarray}
where $L_{\theta}$ is the symmetric logarithmic derivative (SLD). $L_{\theta}$ is given as
\begin{eqnarray}
    \frac{\partial}{\partial\theta}\rho_\theta=\frac{1}{2}(\rho_{\theta}L_{\theta}+L_{\theta}\rho_{\theta}).
\end{eqnarray}

If $\rho_\theta$ is a pure state $\rho_\theta=\ketbra{\psi_{\theta}}{\psi_{\theta}}$, $F_{\text{Q}}$ can be computed as below:
\begin{equation}
    F_{\text{Q}}=4\text{Re}(\braket{\partial_\theta{\psi_{\theta}|\partial_{\theta}\psi_{\theta}}}-|\braket{\partial_{\theta}\psi_{\theta}|\psi_{\theta}}|^2).
\end{equation}
where $\ket{\partial{\theta\psi_{\theta}}}=\frac{\partial}{\partial\theta}\ket{\psi_{\theta}}$. In multiparameter scenario, the QFI is replaced by the  Fisher information matrix (QFIM) \cite{HL, FIM}:

\begin{equation}
    (\bm{F}_\text{Q}[\rho_{\theta}])_{kl}=4\text{Re}(\braket{\partial_{\theta_{k}}\psi_{\theta}|\partial_{\theta_l}\psi_{\theta}}-\braket{\partial_{\theta_k}\psi_{\theta}|\psi_{\theta}}\braket{\psi_{\theta}|\partial_{\theta_l}\psi_{\theta}})\label{definition_of_QFIp}
\end{equation}
where $\ket{\partial_{\theta_k}\psi_{\theta}}=\frac{\partial}{\partial\theta_k}\ket{\psi_{\theta}}$. 

Let us note that the optimal measurement in the multi-parameter estimation that satisfies $F_{\text{C}}=F_{\text{Q}}$ does not always exist.

To calculate $F_{\text{Q}}$ in the direct transmission protocol, we only use the pure GHZ state, because (ignoring the detector dark counts) the case when it is generated can be post-selected based on all photons being detected. So, the parametrized GHZ state simply becomes:
\begin{equation}
\rho_{\text{D},{\theta}}=U_{\theta}\ketbra{\text{GHZ}}{\text{GHZ}}U^{\dagger}_{\theta},\label{ghzdp}
\end{equation}
where $\ket{\text{GHZ}}$ is described in Eq.~(\ref{ghz_polari}). To calculate the CFIM, we choose the $\sigma_x$ basis measurement in Eq.~(\ref{sigmaxmeasur}).
Following (\ref{definition_of_QFIp}), the QFIM of this state reads
\begin{equation}
        (\bm{F}_\text{Q}[\rho_{\text{D},{\theta}}])_{kl}=1,
\end{equation}
if $k=l$ or both $k$ and $l$ are even or odd; or
\begin{equation}
        (\bm{F}_\text{Q}[\rho_{\text{D},{\theta}}])_{kl}=-1, 
\end{equation}
if $k$ is even and $l$ is odd or $k$ is odd and $l$ is even. Finally, we get the QFIM,
\begin{equation}
    \bm{F}_\text{Q}[\rho_{\text{D},{\theta}}]=
    \begin{bmatrix}
        1 & -1 & 1 & -1\\
        -1 & 1 & -1 & 1\\
        1 & -1 & 1 & -1\\
        -1 & 1 & -1 & 1
    \end{bmatrix}.\label{qfimd}
\end{equation}
By diagonalizing this matrix, we get only one nonzero eigenvalue $\lambda$ and the corresponding eigenvector $\ket{\lambda}$ as follows:
\begin{equation}
    \lambda=4,\qquad \ket{\lambda}=\{1,-1,1,-1\}. \label{evecval_dd}
\end{equation}
Hence,
\begin{equation}
    F_{\text{Q}}[\rho_{\text{D}, \theta}]=\lambda=4, \qquad \Delta^2\theta\geq\frac{1}{4N}.\label{fqd}
\end{equation}
We note that the derived QCRB (\ref{fqd}) indicates the Heisenberg scaling in multiparameter estimation for $M=4$. In the classical case using the separable state $\ket{\varphi}=\otimes_{i=0}^{3}\frac{1}{\sqrt{2}}(\ket{0}+\ket{1})_i$, the QCRB can be calculated as $\Delta^2\theta=\Sigma_{i=0}^{3}\Delta^2(\theta_i)\geq\frac{4}{N}$, corresponding to the SQL.

When we project state (\ref{ghzdp}) onto the $\sigma_x$ basis, we can obtain two patterns of four coincident detection events depending on $\theta$:
\begin{equation}
    P_1=\frac{1}{16}\left(1+\cos{\theta}\right),\quad P_2=\frac{1}{16}\left(1-\cos{\theta}\right).
\end{equation}
Here $P_1$ corresponds to measurement pattern $(0101)$, and $P_2$ to $(1010)$. 

From Eq.~(\ref{definition_of_CFIM}), the CFIM of (\ref{ghzdp}) is given as
\begin{equation}
    \begin{split}
        (\bm{F}_\text{C}&[\rho_{\text{D},\theta}])_{kl}=8\times\frac{16}{1+\cos{\theta}}\left({\frac{\sin{\theta}}{16}}\right)^2\\
        &+8\times\frac{16}{1-\cos{\theta}}\left(\frac{\sin{\theta}}{16}\right)^2 \\
        &=1,
    \end{split}
\end{equation}
if $k=l$ or both $k$ and $l$ are even or odd; and
\begin{equation}
    \begin{split}
        (\bm{F}_\text{C}&[\rho_{\text{D},{\theta}}])_{kl}=8\times\frac{16}{1+\cos{\theta}}\left(\frac{\mp\sin{\theta}}{16}\right)\left(\frac{\pm\sin{\theta}}{16}\right)\\&+8\times\frac{16}{1-\cos{\theta}}\left(\frac{\pm\sin{\theta}}{16}\right)\left(\frac{\mp\sin{\theta}}{16}\right)\\
        &=-1,
    \end{split}
\end{equation}
if $k$ even and $l$ is odd or $k$ odd and $l$ is even. Hence, 
\begin{equation}
        \bm{F}_\text{C}[\rho_{\text{D},{\theta}}]=
        \begin{bmatrix}
            1 & -1 & 1 & -1\\
            -1 & 1 & -1 & 1\\
            1 & -1 & 1 & -1\\
            -1 & 1 & -1 & 1
        \end{bmatrix}\label{cfimd}
        =\bm{F}_\text{Q}[\rho_{\text{D},{\theta}}]
\end{equation} 
Therefore, $F_\text{Q}[\rho_{\text{D},\theta}]=F_\text{C}[\rho_{\text{D},\theta}]$.

In our scheme, the parametrized state is expressed by (\ref{PGHZstate}). With that state, (\ref{definition_of_QFIp}) becomes:
\begin{equation}
    \hspace{-1pt}\bm{F}_\text{Q}[\rho_{\text{CS},\theta}]\hspace{-2pt}=\hspace{-3pt}\bm{F}_\text{Q}\left[p U_{\theta}\ketbra{\text{GHZ}}{\text{GHZ}}U^{\dagger}_{\theta}\hspace{-2pt}+\hspace{-2pt}\sum_{i}r_{i}\ketbra{\phi_i}{\phi_i}\right].\label{QFICS}
\end{equation}
In this case, the coefficients $p, r_i$ and the noisy terms $\ketbra{\phi_i}{\phi_i}$ are independent of $\theta$, because $p, r_i$ are calculated by the constant $a,b$ and the probability of the photon loss, and $\ketbra{\phi_i}{\phi_i}$ are diagonal matrices. Therefore, we can rewrite Eq.~(\ref{QFICS}) as
\begin{align}
    \bm{F}_\text{Q}[\rho_{\text{CS},{\theta}}] &= p\bm{F}_\text{Q}[U_{\theta}\ketbra{\text{GHZ}}{\text{GHZ}}_4U_{\theta}^\dagger] \nonumber \\
         &=p\left[
         \begin{array}{cccc}
             1 & -1 & 1 & -1 \\
             -1 & 1 & -1 & 1 \\
             1 & -1 & 1 & -1 \\
             -1 & 1 & -1 & 1 
         \end{array}
         \right]
         \label{qfimcs_mat}
\end{align}
which implies
\begin{equation}
    F_{\text{Q}}[\rho_{\text{CS},{\theta}}] = 4p.
\end{equation}
To calculate the CFIM of our scheme considering the measurement (\ref{displace_and_pc}), we get the probabilities of several possible coincidence detection events. In contrast to the direct transmission, we can get nontrivial patterns. 
The results are discussed in Section III. 

\begin{figure*}[htbt]
    \centering
    \includegraphics[width=1\linewidth]{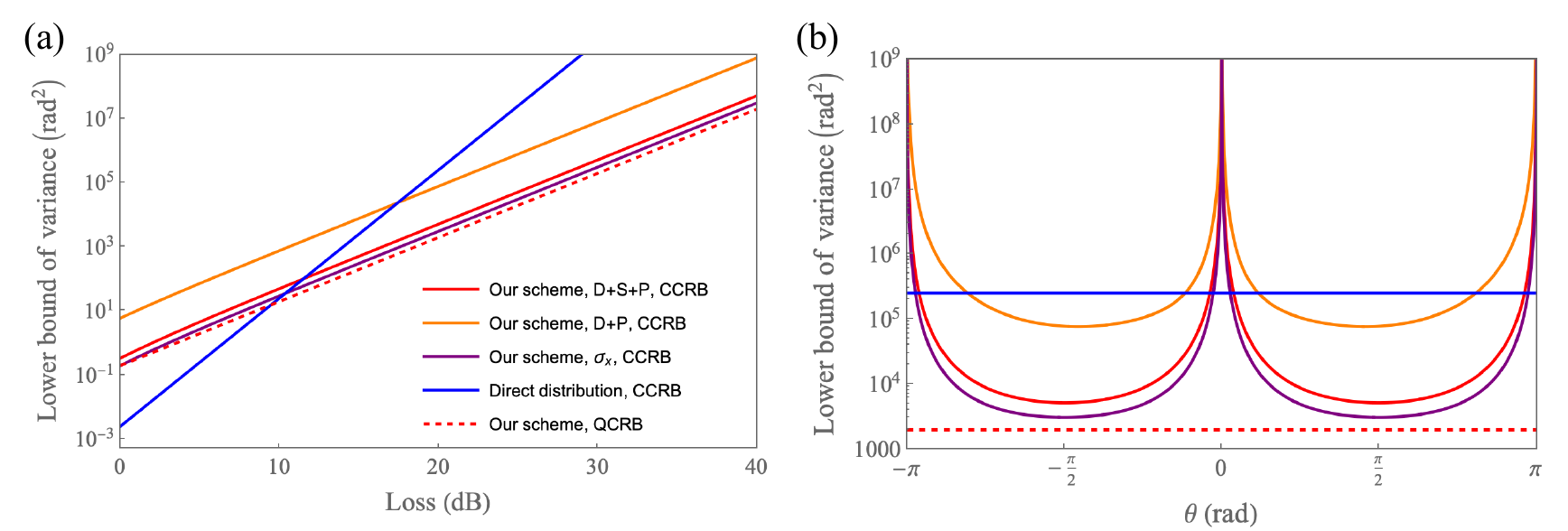}
    \caption{Lower bound of variance derived by QCRB and CCRB. QCRB of our protocol is shown as the red dotted line. Our protocol measured by squeezing, displacement, and photon counting is shown as the red line denoted by D+S+P. Our protocol in which the qubits are projected onto the $\sigma_x$ basis is shown as the purple line. The orange line denoted by D+P shows the CCRB of our scheme measured by displacement and photon counting. The blue line shows the CCRB of Direct distribution. (a) The lower bound of variance versus distribution loss when $\theta=\frac{\pi}{2}$. (b) The lower bound of variance versus the sensed phase $\theta=\theta_0-\theta_1+\theta_2-\theta_3$ when the distribution loss is 20 dB. 
    }
    \label{5}
\end{figure*}

\section{\label{sec:level2-1}Results}
We calculate the QCRB and the CCRB for the scenario with 4 sensing stations and 4 parameters. 
Assuming the fiber loss $0.2$ db/km and the distance $L$, the corresponding transmittance is $\eta = 10^{-0.2L/10}$. In Fig.~\ref{5}, we plot the variance as a function of the total loss in dB $(-10\log_{10}\eta)$. 
We assume the number of repetitions $N=100$, and optimize the displacement parameter $\alpha$ and the squeezing parameter $\xi$ to get the minimum CCRB. We have chosen the parameters $a$ and $b$ in (\ref{bell1}) as follows: $a=\sqrt{0.8}$ and $b=\sqrt{0.2}$ which give a reasonable fidelity and rate of the generated GHZ state \cite{shimizughz}. 
Figure~\ref{5}, shows the results for the parameter $\theta=\theta_0-\theta_1+\theta_2-\theta_3$, which is estimated if the detection registers the generation of $\ket{\Psi}=\frac{1}{\sqrt{2}}(\ket{1010}-\ket{0101})$. We compare the performance of our scheme with the conventional method---direct transmission. The set of probability function $\bm{P}$ is constructed based on four-coincidence events of the $\sigma_x$ measurements in the local stations.

For precision evaluation including the distribution loss, we assume the number of valid measurements $N^{'}$ as:
\begin{equation}
    N^{'}=N\times P_{\text{suc}}
\end{equation}
where $P_{\rm suc}$ is the success probability of GHZ state distribution. The details are shown in Appendix \ref{A}. 

As we mentioned in Section \ref{sec:level2}, the QCRB and the CCRB in our scheme are calculated using displacement, squeezing, and photon detection. For comparison, in our scenario, we also calculate the quantities in two cases; (1)~considering $\sigma_x$ basis measurement as the ideal case, and (2)~using displacement and photon detection. The second situation is easier to implement than the method with squeezing.

Figure \ref{5}(a) shows that our scheme has an advantage in the scaling of the variance. In addition, our measurement setup achieves almost same performance as the case using $\sigma_x$ basis measurement. However, the CCRB of our scheme does not reach the QCRB. In the ideal case of our scheme, using $\sigma_x$ basis measurement, the CCRB reaches the QCRB when there is no loss but it does not reach it in a presence of loss. It means that $\sigma_x$ basis measurement is not optimal for noisy states. 

On the other hand, the strategy with only displacement, shown as the orange line in Fig.~\ref{5}, has the larger variance than the strategy with squeezing, displacement and photon detection. But comparing with the direct distribution case, it still has an advantage in the scaling of the variance and achieves smaller variance than direct distribution when the loss is high.

Figures \ref{5}(b) shows that the variance depends on the estimation parameters. Our scheme provides more precise estimation than the alternative method. However, as in Fig.~\ref{5}(a), the CCRB does not reach the QCRB. Additionally, for specific phases the variance diverges. This phase-dependence comes from the noisy terms including the quantum state. The details regarding this fact are provided in Appendix \ref{B}.

\section{Private quantum network sensing with the additional phase}

In this section, we introduce the notion of privacy in the quantum network sensing proposed in Ref.~\cite{PrivacyCV}. We will show that our protocol satisfies the condition of privacy. We also extend this work to the case in which only the central station can estimate the target linear combination of parameters $\theta_t=\theta_0-\theta_1+\theta_2-\theta_3$ privately, while the other nodes cannot estimate it. In general, the estimated parameter $f(\bm{\theta})$ is described as $f(\bm{\theta})=\bm{\nu}\cdot\bm{\theta}$, where $\bm{\nu}$ is the weight vector, $\{\nu_0,\nu_1,...,\nu_{M-1}\}$, and $\bm{\theta}=\{\theta_0, \theta_1, ..., \theta_{M-1}\}$.

\begin{figure}[htbp]
    \centering
    \includegraphics[width=1\linewidth]{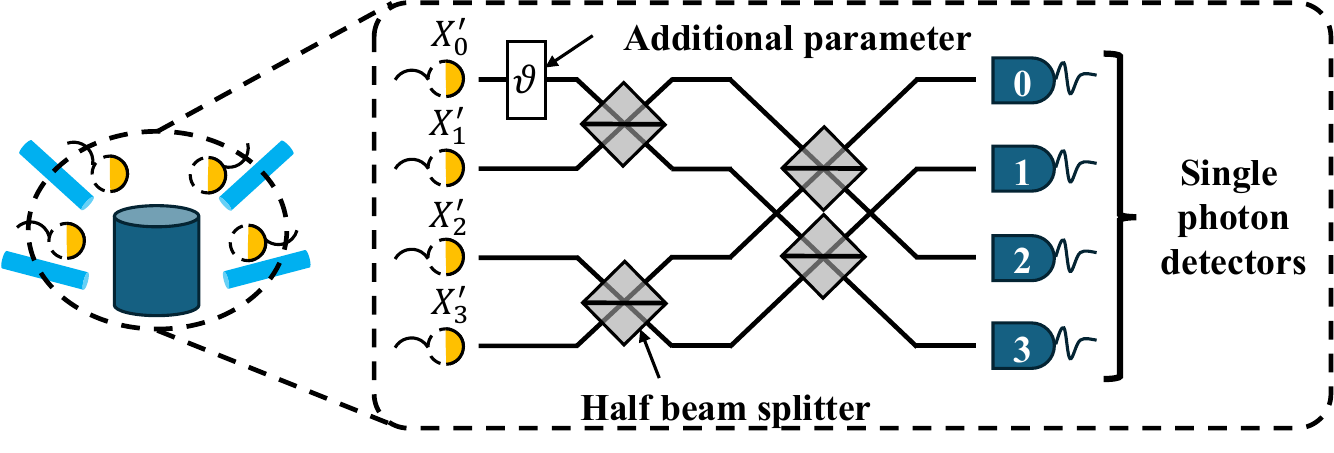}
    \caption{The setting of the central station in the private quantum network sensing scenario. The central station randomly encodes $\vartheta$ before the interference and operates the GHZ state generation and the phase estimation. 
    }
    \label{6}
\end{figure}

Consider the case shown in Fig.~\ref{6}. In this scenario, the central station randomly encodes the additional phase $\vartheta$ on one of transmitted photons before the interference. Therefore, the input states of the interferometer are described as
\begin{equation}
    \begin{split}
        &\ket{\psi}_{X_0X_0^{'}}=a\ket{00}+be^{-i\vartheta}\ket{11}, \\
        &\ket{\psi}_{X_jX_j^{'}}=a\ket{00}+b\ket{11}, j\neq0.
    \end{split}\label{pqns_iqs}
\end{equation}
Thus, the generated GHZ state is given as
\begin{equation}
    \ket{\text{GHZ}_{\vartheta}}=\frac{1}{\sqrt{2}}\left(e^{-i\vartheta}\ket{1010}_{X_0X_1X_2X_3}-\ket{0101}_{X_0X_1X_2X_3}\right). 
    \label{GHZvt}
\end{equation}
The detail of this generation is given in Appendix.~C. In this scenario, we assume that the central station to estimate $\theta_t$ uses the GHZ state from Eq.~\eqref{GHZvt}.

\subsection{The notion of privacy}
Let us define an "unobservable" and the indicator $P(\bm{F}_{\text{Q}},\bm{\nu})$ of privacy based on the QFI, proposed in Ref.~\cite{PrivacyCV}. These objects allow us to quantify the level of privacy, which is classified into two levels; 1) Strong (complete) privacy, which means that only the linear combination $f(\bm{\theta})$ of the local parameters is accessible and neither the individual parameters nor any linear combinations of the parameters except $f(\bm{\theta})$ can be known. 2) Weak privacy, which means that some extra collective information, such as other combinations of parameters, could be learned, but the individual phases remain hidden.

\begin{dfn} 
    [Unobservable and private parameter \cite{PrivacyCV}]
    For $\{\nu_0,\nu_1,...,\nu_{M-1}\} \in\mathbb{R}^{M}$ define the directional derivative: \\
    \begin{equation}
        \partial_{\bm{\nu}}\rho_{\bm{\theta}}:=\sum_{\mu=0}^{M-1}\nu_\mu\frac{\partial}{\partial\theta_\mu}\rho_{\bm{\theta}}.
    \end{equation} \\
    (i) A vector $\bm{\nu}$ is an unobservable at $\bm{\theta}$ if $\partial_{\bm{\nu}}\rho_{\bm{\theta}}=0$. \\
    (ii) A component $\theta_j$ is private at $\bm{\theta}$ if there exists an unobservable $\bm{\nu}$ with $\nu_j \neq 0$.
\end{dfn}

From (i) of Definition 1, $\partial_{\bm{\nu}}\rho_{\bm{\theta}}=0$ is equivalent to $\abs{\rho_{(\bm{\theta}+t\bm{\nu})}-\rho_{\bm{\theta}}}=\mathcal{O}(t^2)$ as $t\rightarrow 0$. This indicates that shifting the parameter along $\bm{\nu}$ changes $\theta_j$ but does not change the measurement statistics. Thus, $\theta_j$ cannot be fixed in this condition. In consequence of Definition 1 (ii), we can characterize the privacy by the QFI as follows:
\begin{prop}
    [Private parameters]
    A component $\theta_j$ is private at $\bm{\theta}$ iff there exists $\bm{\nu}\in\text{ker}\bm{F}_{\text{Q}}$ with $\nu_j\neq0$.
\end{prop}
As described in Ref.~\cite{PrivacyCV}, $\partial_{\bm{\nu}}\rho_{\bm{\theta}}=0$ is equivalent to $\bm{\nu}\in\text{ker}\bm{F}_{\text{Q}}$ so that the privacy can be defined by the Fisher information shown as Proposition 1. 

Based on Proposition 1, we can define the weak form of privacy as the dimension of $\text{ker}\bm{F}_{\text{Q}}$ while $\text{dim ker}\bm{F}_{\text{Q}}>0$. This condition means that at least one linear combination of parameters is unobservable, but it does not promise that all the individual parameters are hidden. To prove the strong (complete) privacy, we introduce the indicator $P(\bm{F}_{\text{Q}},\bm{\nu})$ described as
\begin{equation}
    P(\bm{F}_{\text{Q}},\bm{\nu})=\frac{\bm{\nu}\bm{F}_{\text{Q}}\bm{\nu}^{\top}}{\Tr[\bm{F}_{\text{Q}}]}.\label{p_ind}
\end{equation}
Equation~(\ref{p_ind}) quantifies the ratio of the information of $\bm{\nu}\cdot\bm{\theta}$. From these indicators, we rewrite the levels of privacy as follows:
\begin{enumerate}
\item Complete privacy, where $P=1$, 
\item Partially private, where $0<P<1$ and $\text{dim ker}\bm{F}_{\text{Q}}>0$, and 
\item Not private, where $0\leq P<1$ and $\text{dim ker}\bm{F}_{\text{Q}}=0$.
\end{enumerate}

\subsection{Privacy in our scheme}
In this section, we present the condition that is necessary to guarantee the required privacy, and  calculate $P(\bm{F}_{\text{Q}},\bm{\nu})$ to show that our scheme satisfies it. 

In our scenario we require complete privacy. This means that the central station can estimate the parameter $\hat{\theta}=\theta_0-\theta_1+\theta_2-\theta_3$ but it cannot estimate any other linear combination of parameters including individual parameters. In addition, the other nodes cannot estimate any parameters including $\hat{\theta}$. In what follows, we show that this requirement is satisfied by the protocol.

First, we calculate the QFIM when the GHZ state in Eq.~\eqref{GHZvt} is used. The parametrized state is given as
\begin{equation}
    U_{\theta}\rho_{\text{CS}^{'}}U_{\theta}^{\dagger}=pU_{\theta}\ketbra{\text{GHZ}_{\vartheta}}{\text{GHZ}_{\vartheta}}U_{\theta}^{\dagger}+\sum_ir_i\ketbra{\phi_i}{\phi_i}
    \label{pqns_param_GHZ}
\end{equation}
where $\rho_{\text{CS}^{'}}=p\ketbra{\text{GHZ}_{\vartheta}}{\text{GHZ}_{\vartheta}}+\sum_ir_i\ketbra{\phi_i}{\phi_i}$. In this scenario, the central station randomly chooses $\vartheta\in[0,2\pi]$. Therefore, $\vartheta$ is constant for the central station and the QFIM of it is described by Eq.~\eqref{qfimcs_mat}.

Next, we prove that the QFIM of Eq.~\eqref{qfimcs_mat} satisfies the condition of complete privacy. The kernel vector of this QFIM is
\begin{equation}
    \text{ker}\bm{F}_{\text{Q,CS}^{'}}=\{1,1,1,1\}.
\end{equation}
Thus, all local parameters are private. The weight vector in our scheme is $\bm{w}=\{1,-1,1,-1\}$. By normalizing $\bm{w}$, $\bm{\nu}$ is $\bm{\nu}=\{\frac{1}{2},-\frac{1}{2},\frac{1}{2},-\frac{1}{2}\}$. The QFIM in our scheme is derived in Eq.~(\ref{qfimcs_mat}). Therefore, $P(\bm{F}_{\text{Q}},\bm{\nu})$ is 
\begin{equation}
    P(\bm{F}_{\text{Q,CS}^{'}},\bm{\nu})=\frac{p\times1}{p\times1}=1.\label{p_ind_os}
\end{equation}
Moreover, the rank of the QFIM in our scheme is one, Eq.~(\ref{evecval_dd}). Thus, the protocol satisfies the condition of complete privacy.

On the other hand, the nodes other than the central station do not have any knowledge about $\vartheta$. Therefore, their quantum states are the states averaged over $\vartheta$, as follows,
\begin{equation}
    \begin{split}
        \rho_{\text{CS}^{'}}^{\text{ave}}&=\int_0^{2\pi}\frac{1}{2\pi}{U}_{\theta}\rho_{\text{CS}^{'}}U_{\theta}^{\dagger}d\vartheta \\
        &=p\int_0^{2\pi}\frac{1}{2\pi}{U}_{\theta}\ketbra{\text{GHZ}}{\text{GHZ}}_{\vartheta}U_{\theta}^{\dagger}d\vartheta+\sum_ir_i\ketbra{\phi_i}{\phi_i}.
        \label{pqns_ave_state}
    \end{split}
\end{equation}
According to $\int_0^{2\pi}\frac{1}{2\pi}e^{i\vartheta}d\vartheta=0$, the state in Eq.~\eqref{pqns_ave_state} has only diagonal terms. Thus, the QFIM of the nodes other than the central station is $\bm{0}$, which indicates that no parameters can be estimated.



\section{\label{sec:level3} Conclusion and outlook}

In this paper, we propose the quantum network sensing protocol using robust GHZ state distribution and calculate the quantum and classical CRB. We consider estimating the linear combination $\theta$ of distributed four parameters in the four users network. We report two main results. First, we show that our protocol achieves lower variance of estimation than the reference conventional method in the regime of high loss. However, compared with the conventional method, our protocol has a restriction in terms of measurement, because the distributed GHZ state in our scheme is encoded in the photon-number basis. Nevertheless, our measurement method using displacement, squeezing and photon detection achieves almost the same lower bound as in case of the optimal measurement of the conventional case. Second, by considering another setup in which the central station encodes the additional parameter, our protocol can measure $\theta_t$ without not only revealing the individual parameters but also $\theta_t$ to the other network nodes. 

To approach QCRB, we may consider possible extensions of the scheme including optimization of the measurement and alternative choices of the probe states. Regarding the measurement optimization, the $\sigma_x$ basis measurement is known as the measurement which can achieve QCRB in the ideal situation, but when the noisy state appears, it remains suboptimal. Furthermore, we cannot estimate certain phases because the variance becomes infinite, as shown in Fig.~\ref{5}(b). Moreover, for CCRB to better approach QCRB we need to construct the measurement scheme which is robust against noise. 

Entangled states different than GHZ can also be used in the quantum network sensing context.  
Recent research \cite{lossghz} suggests that symmetric Dicke states are robust against noise and enable more precise estimation than GHZ states. The protocol we consider here is flexible and can be adapted to generate not only GHZ states, but also Dicke states depending on the measurement pattern \cite{wojciechWDicke}. For quantum network sensing applications, it is necessary to investigate the optimal quantum setups further.

In this work, we assume that the reference conventional method---direct transmission---can generate and distribute pure GHZ state deterministically so that CCRB achieves the QCRB. However, in realistic situations, this method sometimes generates noisy GHZ states or fails to generate. In this work we idealize the reference method to search for the worst case advantages of our scheme. Indeed, we calculate QCRB and CCRB including the noise in the generation step and the success probability of GHZ state generation in our method. As expected and shown in our results and \cite{lossghz}, the variance increases if the used states are noisy. For more realistic comparison it is reasonable to calculate CCRB including the noise and generation rate also in the reference method.

Also, as a proof of principle, we consider a 4-parameter problem. Nevertheless, our scheme can be extended beyond that number. The direction of further studies would be to look for the Heisenberg scaling against the number of sensing stations $M$. Increasing this number could be essential in future applications. It is known that, due to fiber loss, the success probability of entangled state sharing is exponentially decreasing when $M$ increases. However, as shown in the Results section, our scheme is relatively robust against fiber loss as the scaling of the success rate is $\eta^{\frac{M}{2}}$. These features support our confidence that the scheme proposed here is effective for future applications.

Our protocol provides a setup not only for precise multiparameter estimation, but also for private sensing. In this work, we assume that users follow the measurement procedure. However, if they decide to sabotage, 
the quantum state generation and swap the quantum state prepared by them, the privacy defined in our paper is not enough. To protect against such attacks, we need an additional process, such as quantum state verification. We note that our protocol can distribute the GHZ states efficiently. It is suitable to apply quantum state verification since such verification process requires many quantum states.

In terms of private network sensing, it is interesting to apply the anonymous sensing protocol. Our scheme allows for estimation of certain linear combinations of the local phases (those corresponding to the nonzero eigenvectors of the QFIM and the detection patterns in Table II), and by encoding a known local phase at node 0, one can privately compute a particular combination. This fact clearly shows the potential for the applications that require privacy such as voting and the communication of personal information in the medical field.

\begin{acknowledgments}
The authors appreciate Alexssandre de Oliveira J. for productive and helpful discussions. This work was supported by JST Moonshot $R\&D$, Grant No. JPMJMS226C and Grant No. JPMJMS2061, JST CRONOS, Grant No. JPMJCS24N6, JST SPRING, Grant No. JPMJSP2123, JST ASPIRE,
Grant No. JPMJAP2427, JST COI-NEXT Grant No. JPMJPF2221 and Program for the
Advancement of Next Generation Research Projects, Keio University.
\end{acknowledgments}



\appendix

\section{Success probability of GHZ state generation}\label{A}

Here we show the calculation of the success probability of four-partite GHZ state generation. As we mentioned before, distribution process in the direct transmission method, which is our reference, records a success if all photons arrive at the end stations. Therefore, the success probability is $\eta^4$.

In our scheme, GHZ states are generated when we get the target detection pattern from single photon detectors. According to Ref.~\cite{shimizughz}, this probability is 
\begin{equation}
    \begin{split}
        P_{\text{CS}}&=2\sum_{m=0}^{\frac{N}{2}}\eta^{\frac{N}{2}}\left(1-\eta\right)^{m}\left(\frac{1}{2}\right)^{\frac{3}{2}N-4}\\
        &\times a^{N-2m}b^{N+2m}{}_\frac{N}{2}C_m\\
        &=\left(\frac{1}{2}\right)^{\frac{3}{2}N-5}\eta^{\frac{N}{2}}b^N\left[a^2+b^2(1-\eta)\right]^{\frac{N}{2}}.\label{spcsghz}
    \end{split}
\end{equation}
where $m$ is the number of photons which is lost in the fiber, and $N$ is the number of the nodes. Equation.~(\ref{spcsghz}) takes into account the situation when photon loss can occur. Specifically, we get the target detection pattern, with probabilities
\begin{equation}
    \begin{split}
        &P_{\text{CS},0}=2\eta^{\frac{N}{2}}\left(\frac{1}{2}\right)^{\frac{3}{2}N-4}a^{N}b^{N},\\
        &P_{\text{CS},1}=2\eta^{\frac{N}{2}}(1-\eta)\left(\frac{1}{2}\right)^{\frac{3}{2}N-4}a^{N-2}b^{N+2}{}_\frac{N}{2}C_1,\\
        &P_{\text{CS},2}=2\eta^{\frac{N}{2}}(1-\eta)^{2}\left(\frac{1}{2}\right)^{\frac{3}{2}N-4}a^{N-4}b^{N+4}{}_\frac{N}{2}C_2,\label{spcsghzr}
    \end{split}
\end{equation}
with the number of photons lost indicated by the subscript. From Eq.(\ref{spcsghz}) and Eq.(\ref{spcsghzr}), the rates-parameters of the noisy GHZ state taken into account in the paper $p, r_i$ are given by
\begin{equation}
    p=\frac{P_{\text{CS},0}}{P_{\text{CS}}},\quad r_1 = \frac{P_{\text{CS},1}}{P_{\text{CS}}},\quad r_2 = \frac{P_{\text{CS},2}}{P_{\text{CS}}}.
\end{equation}

\section{The divergence range in the result}\label{B}
We mentioned that the lower bound of the variances in our results diverge in the specific range. The reason why such divergent ranges appear is the noise terms, including the quantum state and detection probability. Our protocol can generate the mixed GHZ state described in Eq.(\ref{PGHZstate}) and the undesirable terms are expressed by
\begin{equation}
    \begin{split}
        \ketbra{\psi_1}{\psi_1}=\frac{1}{4}(&\ketbra{0111}{0111}+\ketbra{1011}{1011}\\
        &+\ketbra{1101}{1101}+\ketbra{1110}{1110}),
        \label{lt1}
    \end{split}
\end{equation}
\begin{equation}
    \ketbra{\psi_2}{\psi_2}=\ketbra{1111}{1111}.
    \label{lt2}
\end{equation}
For simplicity, we consider projecting onto the $\sigma_x$ basis. From Eq.~\eqref{PGHZstate}, \eqref{lt1}, and \eqref{lt2}, we can get two types of detection probabilities in 16 patterns:
\begin{equation}
    P_1=p\frac{1+\cos\theta}{16}+\frac{r_1+r_2}{16}=\frac{1+p\cos\theta}{16}
\end{equation}
\begin{equation}
    P_2=p\frac{1-\cos\theta}{16}+\frac{r_1+r_2}{16}=\frac{1-p\cos\theta}{16}.
\end{equation}

Thus, the elements of CFIM are expressed as\\
\begin{equation}
    \begin{split}
        (\bm{F}_\text{C}&[\rho_{\text{CS},\theta}])_{kl}
        \\
        &=8p^2\sin^2{\theta}\left[\frac{(1-p\cos{\theta})+(1+p\cos{\theta})}{16(1-p^2\cos^2{\theta})}\right]\\
        &=\frac{p^2\sin^2{\theta}}{1-p^2\cos^2{\theta}},
    \end{split}
\end{equation}
if $k=l$ or both $k$ and $l$ are even or odd; and
\begin{equation}
    \begin{split}
        (\bm{F}_\text{C}&[\rho_{\text{CS},{\theta}}])_{kl}
        \\
        &=-8p^2\sin^2{\theta}\left[\frac{(1-p\cos{\theta})+(1+p\cos{\theta})}{16(1-p^2\cos^2{\theta})}\right]\\
        &=\frac{-p^2\sin^2{\theta}}{1-p^2\cos^2{\theta}},
    \end{split}
\end{equation}
if $k$ even and $l$ is odd or $k$ odd and $l$ is even. In the pure state model, all elements in CFIM are independent on $\theta$, but the above equations show that the CFIM in the mixed GHZ states has the phase-dependence. If $\sin{\theta}$ becomes zero, the amount of Fisher information is also zero. That is the reason why our protocol has the phase-dependence which comes from the loss term.

\section{The inference in central station with $\vartheta$}
Here, we assume the distribution without losses. From Eq.~\eqref{pqns_iqs}, the distributed state is given as 
\begin{equation}
    \ket{\Psi}=(a\ket{00}+be^{-i\vartheta}\ket{11})\otimes\left[(a\ket{00}+b\ket{11})^{\otimes3}\right].\label{pqns_prod}
\end{equation}
For simplicity, we omit the remained states, $X_0,X_1,X_2,X_3$, and rewrite Eq.~\eqref{pqns_prod} with the creation operators $\hat{a}^{\dagger}_j$ as follows:
\begin{align}
    \ket{\Psi^{'}}&=\left[(a+b^{'}\hat{a}_0^{\dagger})\times\prod_{j=1}^{3}(a+b\hat{a}_j^{\dagger})\right]\ket{0000}_{X_0^{'}X_1^{'}X_2^{'}X_3^{'}} \nonumber \\
        &=[a^4+a^3(b^{'}\hat{a}_0^{\dagger}+b\hat{a}_1^{\dagger}+b\hat{a}_2^{\dagger}+b\hat{a}_3^{\dagger}) \nonumber \\
        &\hspace{13pt}+a^2(b^{'}b\hat{a}^{\dagger}_{0}\hat{a}^{\dagger}_{1}+b^{'}b\hat{a}^{\dagger}_{0}\hat{a}^{\dagger}_{2}+b^{'}b\hat{a}^{\dagger}_{0}\hat{a}^{\dagger}_{3}) \nonumber \\
        &\hspace{13pt}+a^2(b^2\hat{a}^{\dagger}_{1}\hat{a}^{\dagger}_{2}+b^2\hat{a}^{\dagger}_{1}\hat{a}^{\dagger}_{3}+b^2\hat{a}^{\dagger}_{2}\hat{a}^{\dagger}_{3}) \nonumber \\
        &\hspace{13pt}+a(b^{'}b^2\hat{a}^{\dagger}_{0}\hat{a}^{\dagger}_{1}\hat{a}^{\dagger}_{2}+b^{'}b^2\hat{a}^{\dagger}_{0}\hat{a}^{\dagger}_{1}\hat{a}^{\dagger}_{3}) \nonumber \\
        &\hspace{13pt}+a(b^{'}b^2\hat{a}^{\dagger}_{0}\hat{a}^{\dagger}_{2}\hat{a}^{\dagger}_{3}+b^3\hat{a}^{\dagger}_{1}\hat{a}^{\dagger}_{2}\hat{a}^{\dagger}_{3}) \nonumber \\
        &\hspace{13pt}+b^{'}b^3\hat{a}^{\dagger}_0\hat{a}^{\dagger}_1\hat{a}^{\dagger}_2\hat{a}^{\dagger}_3]\ket{0000}.
        \label{pqns_prod_exp}
\end{align}
According to \cite{shimizughz}, the interference in the central station is described as
\begin{equation}
    \left[
    \begin{array}{c}
        \hat{b}^{\dagger}_0 \\
        \hat{b}^{\dagger}_1 \\
        \hat{b}^{\dagger}_2 \\
        \hat{b}^{\dagger}_3
    \end{array}\right]
    =U_{\text{BSN}}\left[
    \begin{array}{c}
        \hat{a}^{\dagger}_0 \\
        \hat{a}^{\dagger}_1 \\
        \hat{a}^{\dagger}_2 \\
        \hat{a}^{\dagger}_3
    \end{array}\right],
\end{equation}
\begin{equation}
    U_{\text{BSN}}=\frac{1}{2}\left[
    \begin{array}{cccc}
        1 & 1 & 1 & 1 \\
        1 & -1 & 1 & -1 \\
        1 & 1 & -1 & -1 \\
        1 & -1 & -1 & 1 
    \end{array}\right]
\end{equation}
where $\hat{b}_j^{\dagger}$ is the creation operator of output modes. Now, we are interested in the two-photon detection event. So we only consider the state with two photons. Therefore, the interfered two-photon state is given as
\begin{widetext}
\begin{equation}
    \begin{split}
        &a^2(b^{'}b\hat{a}^{\dagger}_{0}\hat{a}^{\dagger}_{1}+b^{'}b\hat{a}^{\dagger}_{0}\hat{a}^{\dagger}_{2}+b^{'}b\hat{a}^{\dagger}_{0}\hat{a}^{\dagger}_{3}+b^2\hat{a}^{\dagger}_{1}\hat{a}^{\dagger}_{2}+b^2\hat{a}^{\dagger}_{1}\hat{a}^{\dagger}_{3}+b^2\hat{a}^{\dagger}_{2}\hat{a}^{\dagger}_{3}) \\
        &\rightarrow\frac{1}{4}a^2[b^{'}b(\hat{b}^{\dagger^2}_{0}-\hat{b}^{\dagger^2}_{1}+\hat{b}^{\dagger^2}_{2}-\hat{b}^{\dagger^2}_{3}+2\hat{b}^{\dagger}_{0}\hat{b}^{\dagger}_{2}-2\hat{b}^{\dagger}_{1}\hat{b}^{\dagger}_{3})+b^{'}b(\hat{b}^{\dagger^2}_{0}+\hat{b}^{\dagger^2}_{1}-\hat{b}^{\dagger^2}_{2}-\hat{b}^{\dagger^2}_{3}+2\hat{b}^{\dagger}_{0}\hat{b}^{\dagger}_{1}-2\hat{b}^{\dagger}_{2}\hat{b}^{\dagger}_{3}) \\
        &\hspace{35pt}+b^{'}b(\hat{b}^{\dagger^2}_{0}-\hat{b}^{\dagger^2}_{1}-\hat{b}^{\dagger^2}_{2}+\hat{b}^{\dagger^2}_{3}+2\hat{b}^{\dagger}_{0}\hat{b}^{\dagger}_{3}-2\hat{b}^{\dagger}_{1}\hat{b}^{\dagger}_{2})+b^2(\hat{b}^{\dagger^2}_{0}-\hat{b}^{\dagger^2}_{1}-\hat{b}^{\dagger^2}_{2}+\hat{b}^{\dagger^2}_{3}-2\hat{b}^{\dagger}_{0}\hat{b}^{\dagger}_{3}-2\hat{b}^{\dagger}_{1}\hat{b}^{\dagger}_{2}) \\
        &\hspace{35pt}+b^2(\hat{b}^{\dagger^2}_{0}+\hat{b}^{\dagger^2}_{1}-\hat{b}^{\dagger^2}_{2}-\hat{b}^{\dagger^2}_{3}-2\hat{b}^{\dagger}_{0}\hat{b}^{\dagger}_{1}-2\hat{b}^{\dagger}_{2}\hat{b}^{\dagger}_{3})+b^2(\hat{b}^{\dagger^2}_{0}-\hat{b}^{\dagger^2}_{1}+\hat{b}^{\dagger^2}_{2}-\hat{b}^{\dagger^2}_{3}-2\hat{b}^{\dagger}_{0}\hat{b}^{\dagger}_{2}+2\hat{b}^{\dagger}_{1}\hat{b}^{\dagger}_{3}).
    \end{split}\label{pqns_tps}
\end{equation}
We rewrite Eq.~\eqref{pqns_tps} with the creation operators of the remained modes in local nodes, $\hat{a}^{\dagger}_{j^{'}}$, as follows:
\begin{equation}
    \begin{split}
        &\frac{1}{4}a^2[b^{'}b\hat{a}^{\dagger}_{0^{'}}\hat{a}^{\dagger}_{1^{'}}(\hat{b}^{\dagger^2}_{0}-\hat{b}^{\dagger^2}_{1}+\hat{b}^{\dagger^2}_{2}-\hat{b}^{\dagger^2}_{3}+2\hat{b}^{\dagger}_{0}\hat{b}^{\dagger}_{2}-2\hat{b}^{\dagger}_{1}\hat{b}^{\dagger}_{3})+b^{'}b\hat{a}^{\dagger}_{0^{'}}\hat{a}^{\dagger}_{2^{'}}(\hat{b}^{\dagger^2}_{0}+\hat{b}^{\dagger^2}_{1}-\hat{b}^{\dagger^2}_{2}-\hat{b}^{\dagger^2}_{3}+2\hat{b}^{\dagger}_{0}\hat{b}^{\dagger}_{1}-2\hat{b}^{\dagger}_{2}\hat{b}^{\dagger}_{3}) \\
        &\hspace{20pt}+b^{'}b\hat{a}^{\dagger}_{0^{'}}\hat{a}^{\dagger}_{3^{'}}(\hat{b}^{\dagger^2}_{0}-\hat{b}^{\dagger^2}_{1}-\hat{b}^{\dagger^2}_{2}+\hat{b}^{\dagger^2}_{3}+2\hat{b}^{\dagger}_{0}\hat{b}^{\dagger}_{3}-2\hat{b}^{\dagger}_{1}\hat{b}^{\dagger}_{2})+b^2\hat{a}^{\dagger}_{1^{'}}\hat{a}^{\dagger}_{2^{'}}(\hat{b}^{\dagger^2}_{0}-\hat{b}^{\dagger^2}_{1}-\hat{b}^{\dagger^2}_{2}+\hat{b}^{\dagger^2}_{3}-2\hat{b}^{\dagger}_{0}\hat{b}^{\dagger}_{3}-2\hat{b}^{\dagger}_{1}\hat{b}^{\dagger}_{2}) \\
        &\hspace{20pt}+b^2\hat{a}^{\dagger}_{1^{'}}\hat{a}^{\dagger}_{3^{'}}(\hat{b}^{\dagger^2}_{0}+\hat{b}^{\dagger^2}_{1}-\hat{b}^{\dagger^2}_{2}-\hat{b}^{\dagger^2}_{3}-2\hat{b}^{\dagger}_{0}\hat{b}^{\dagger}_{1}-2\hat{b}^{\dagger}_{2}\hat{b}^{\dagger}_{3})+b^2\hat{a}^{\dagger}_{2^{'}}\hat{a}^{\dagger}_{3^{'}}(\hat{b}^{\dagger^2}_{0}-\hat{b}^{\dagger^2}_{1}+\hat{b}^{\dagger^2}_{2}-\hat{b}^{\dagger^2}_{3}-2\hat{b}^{\dagger}_{0}\hat{b}^{\dagger}_{2}+2\hat{b}^{\dagger}_{1}\hat{b}^{\dagger}_{3})].
        \label{pqns_term_pat}
    \end{split}
\end{equation}
\end{widetext}
From Eq.~\eqref{pqns_term_pat}, if two photons are detected in mode $0,1$, the generated state is described as
\begin{equation}
    \ket{\Phi}=\frac{1}{\mathcal{N}}\left[\frac{1}{2}a^2b^{'}b\hat{a}^{\dagger}_{0^{'}}\hat{a}^{\dagger}_{2^{'}}-\frac{1}{2}a^2b^2\hat{a}^{\dagger}_{1^{'}}\hat{a}^{\dagger}_{3^{'}}\right]\ket{0000}_{X^{'}_{0}X^{'}_{1}X^{'}_{2}X^{'}_{3}}
\end{equation}
where $\mathcal{N}$ is the normalization factor given as
\begin{equation}
    \mathcal{N}=\sqrt{\abs{\frac{1}{2}a^2b^{'}b}^2+\abs{\frac{1}{2}a^2b^2}^2}=\frac{a^2b^2}{\sqrt{2}}.
\end{equation}
Therefore, the generated state is
\begin{equation}
    \ket{\Phi}=\frac{1}{\sqrt{2}}(e^{-\vartheta}\ket{1010}-\ket{0101}).
\end{equation}

\nocite{*}
\vspace{2pt}
\bibliography{apssamp}

\end{document}